\documentclass[aps,prd]{revtex4}
\usepackage{subfigure,graphics} 
\usepackage{flafter} 
\begin{document}

\title{
Gaugino condensation in an improved heterotic $M$-theory}
\author{Nasr Ahmed}
\email{nasr.ahmed@ncl.ac.uk}
\author{Ian G. Moss}
\email{ian.moss@ncl.ac.uk}
\affiliation{School of Mathematics and Statistics, Newcastle University, NE1 7RU, UK}

\date{\today}


\begin{abstract}

Gaugino condensation is discussed in the context of a consistent new version of low energy heterotic
$M$-theory. The four dimensional reduction of the theory is described, based on simple boson and
fermion backgrounds. This is generalised to include gaugino condenstates and various background
fluxes, some with non-trivial topology.  It is found that condensate and quantised flux
contributions to the four-dimensional superpotential contain no corrections due to the warping of
the higher dimensional metric.
\end{abstract}
\pacs{PACS number(s): }

\maketitle
\section{introduction}

Horava and Witten \cite{Horava:1995qa,Horava:1996ma} proposed some time ago that the low energy
limit of strongly coupled heterotic string theory could be formulated as eleven dimensional
supergravity on a manifold with boundary. Although the theory has received less attention recently
than type IIB superstring theory, it nevertheless possesses advantages over other string theories as
a starting point for particle phenomenology \cite{Witten:1996mz,banks96}.

The original formulation of Horava and Witten contained some serious flaws which where corrected
recently using a new formulation of supergravity on manifolds with boundary 
\cite{Moss:2003bk,Moss:2004ck,Moss:2005zw}. The most serious problem affecting the model was that it
was expressed as a series in factor $\kappa_{11}{}^{2/3}$ 
multiplying the matter action, which worked
well at leading and next-to-leading order but became ill-defined thereafter. This problem was
resolved by a simple modification to the boundary conditions resulting in a low energy theory which
is supersymmetric to all orders in $\kappa_{11}{}^{2/3}$.

It is necessary to revisit important issues such as reduction to four dimensions and moduli
stabilisation, where there has been much progress recently 
\cite{Buchbinder:2003pi,Becker:2004gw,Curio:2005ew,deCarlos:2005kh,
Braun:2006th,Serone:2007sv,Correia:2007sv,Gray:2007qy}, 
in the light of this new formulation of
low-energy heterotic $M$-theory. In this paper we shall focus on the effects of gaugino
condensation, since this is the quantity which is most affected by the new boundary conditions.

In the usual reductions, six of the internal dimensions lie on a Calabi-Yau three-fold and one
internal dimension stretches between the two boundaries. Often, five-branes are included which run
parallel to the boundaries. The reductions have separation moduli between the boundaries or five
branes and Calabi-Yau moduli, which naturally split into two families depending on the type of
harmonic form: $(1,1)$ and $(2,1)$. A variety of stabilisation mechanisms have been proposed:
\begin{itemize}
\item  internal fluxes, which might fix the $(2,1)$ moduli by analogy with the moduli stabilisation
used in type IIB string theory \cite{Kachru:2003aw}
\item branes stretching between the boundaries or the five-branes 
\cite{Buchbinder:2003pi,Braun:2006th}
\item gaugino condensation, which gives a potential depending on the Calabi-Yau volume
\cite{Dine:1985rz,Horava:1996vs,Lukas:1997rb,Gray:2007qy}
\end{itemize}
In addition to the stabilisation mechanisms, there also has to be added some means of ensuring that
the effective cosmological constant is non-negative \cite{Kachru:2003aw,Braun:2006th}.

We shall consider a toy model with the simplest possible set of ingredients, where there is only one
harmonic $(1,1)$ form and no five-branes. We hope to clear up some issues, such as the contribution
which the warping of the metric makes in flux contributions to the superpotential. We are not
aiming to present a fully phenomenologically accurate description of low energy particle physics.

Before proceding, we shall review some of the ingredients of the improved version of low-energy
heterotic
$M$-theory described in Ref. \cite{Moss:2004ck}. 
The theory is formulated on a manifold ${\cal M}$ with a boundary consisting of two disconnected
components ${\cal M}_1$ and ${\cal M}_2$ with identical topology. The eleven-dimensional part of
the action is conventional for supergravity,
\begin{eqnarray}
S_{SG}=&&{1\over 2 \kappa_{11}^2}\int_{\cal M}\left(-R(\Omega)
-\bar\Psi_I\Gamma^{IJK}D_J(\Omega^*)\Psi_K-\frac1{24}
G_{IJKL}G^{IJKL}\right.
\nonumber\\
&&\left.-\frac{\sqrt{2}}{96}\left(\bar\Psi_I\Gamma^{IJKLMP}\Psi_P
+12\bar\Psi^J\Gamma^{KL}\Psi^M\right)G^*_{JKLM}
-\frac{2\sqrt{2}}{11!}
\epsilon^{I_1\dots I_{11}}(C\wedge G\wedge G)_{I_1\dots I_{11}}\right)dv,
\label{actionsg}
\end{eqnarray}
where $G$ is the abelian field strength and $\Omega$ is the tetrad connection. The combination
$G^*=(G+\hat G)/2$, where hats denote the standardised subtraction of gravitino terms to make a
supercovariant expression. 

The boundary terms which make the action supersymmetric are,
\begin{equation}
S_0={1\over  \kappa_{11}^2}\int_{\cal\partial M}\left(
K\mp\frac14\bar\Psi_A\Gamma^{AB}\Psi_B+
\frac12\bar\Psi_A\Gamma^A\Psi_N\right)dv,
\end{equation}
where $K$ is the extrinsic curvature of the boundary. We shall take the  upper sign on the
boundary component $\partial {\cal M}_1$ and the lower sign on the boundary component 
$\partial {\cal M}_2$. 

There are additional boundary terms with Yang-Mills multiplets, scaled by a parameter $\epsilon$,
\begin{equation}
S_1=-{\epsilon\over \kappa_{11}^2}\int_{\cal \partial M}dv
\left(\frac14\left({\rm tr}F^2-\frac12{\rm tr}R^2\right)+
\frac12{\rm tr}\bar\chi\Gamma^AD_A(\hat\Omega^{**})\chi
+\frac14\bar\Psi_A\Gamma^{BC}\Gamma^A{\rm tr}{F}^*_{BC}\chi\right),
\label{action1}
\end{equation}
where $F^*=(F+\hat F)/2$ and $\Omega^{**}=(\Omega+\Omega^*)/2$. The original formulation of Horava
and Witten contained an extra `$\chi\chi\chi\Psi$' term, but it is not present in the new version.
The formulation given in ref. \cite{Moss:2005zw} was only valid to order $R$, and the extension of
the theory to include the $R^2$ term will be reported elsewhere. 

The specification of the theory is completed by boundary
conditions. For the tangential anti-symmetric tensor components,
\begin{equation}
C_{ABC}=\mp \frac{\sqrt{2}}{12}\epsilon\,\left(\omega^Y_{ABC}-\frac12\omega^L_{ABC}\right)
\mp\frac{\sqrt{2}}{48}\epsilon\,{\rm tr}\bar\chi\Gamma_{ABC}\chi.\label{cbc}
\end{equation}
where $\omega^Y$ and $\omega^L$ are the Yang-Mills and Lorentz chern-simons forms. These boundary
conditions replace the modified Bianchi identity in the old formulation. A suggestion along these
lines was made in the original paper of Horava and Witten \cite{Horava:1996ma}. For the gravitino,
\begin{equation}
\Gamma^{AB}\left(P_\pm+\epsilon\Gamma P_\mp \right)\Psi_A=
\epsilon\left( J_Y{}^A-\frac12J_L{}^A\right),\label{gbc}
\end{equation}
where $P_\pm$ are chiral projectors using the outwart-going normals and
\begin{equation}
\Gamma=\frac1{96}{\rm tr}(\bar\chi\Gamma_{ABC}\chi)\Gamma^{ABC}.
\end{equation}
$J_Y$ is the Yang-Mills supercurrent and $J_L$ is a gravitino analogue of the Yang-Mills
supercurrent.

The resulting theory is supersymmetric {\it to all orders} in the parameter $\epsilon$ when working
to order $R^2$. The gauge, gravity and supergravity anomalies vanish if
\begin{equation}
\epsilon={1\over 4\pi}\left({\kappa_{11}\over4\pi}\right)^{2/3}.
\end{equation}
Further details of the anomaly cancellation, and additional Green-Schwarz terms, can be found in
Ref. \cite{Moss:2005zw}.

\section{background}

The reduction to four dimensions begins with a family of solutions to the field equations which are
homogeneous in four dimensions. In this section we shall adapt the background-field solutions used
orginally by Witten \cite{Witten:1996mz}, 
(further developed in \cite{lukas98,Lukas:1998tt,Lukas:1998ew}),
to the new formulation of low-energy heterotic $M$-theory. These solutions where obtained order by
order in $\epsilon$. We expand the solutions in a similar way, but starting from an action which is
valid to all orders in $\epsilon$ gives better control of the error terms.

The anzatz for the background metric is based on a warped product $M\times S^1/Z_2\times Y$ where
$Y$ is a Calabi-Yau space.  In this metric there are two copies of the
$4-$dimensional manifold $M$, $M_1$ and $M_2$, separated by a distance $l_{11}$.  Ideally, a typical
value for the inverse radius of the Calabi-Yau space would of order the Grand Unification scale
$10^{16}$GeV and the inverse separation would be of order $10^{14}$GeV.

The explicit form of the background metric anzatz which we shall use is
\begin{equation}
ds^2=V^{-2/3}dz^2+V^{-1/3}\eta_{\mu\nu}dx^\mu dx^\nu+
V^{1/3}(\tilde g_{a\bar b}dx^adx^{\bar b}+\tilde g_{\bar ab}dx^{\bar a}dx^b).\label{metric}
\end{equation}
where $\eta_{\mu\nu}$ is the Minkowski metric on $M$, $\tilde g_{a\bar b}$ a fixed metric on $Y$ and 
$V\equiv V(z)$, $z_1\le z\le z_2$. Our background metric anzatz is similar to one used by Curio and
Krause \cite{Curio:2000dw}, except that we use a different coordinate $z$ in the $S_1/Z_2$
direction. 

For simplicity, we shall restrict the class of Calabi-Yau spaces to those with only one harmonic
$(1,1)$ form with components $i\tilde g_{a\bar b}$. In this case the background flux for the
antisymmetrc tensor field depends on only one parameter $\alpha$,
\begin{equation}
G_{ab\bar c\bar d}=\frac13\alpha\left(\tilde g_{a\bar c}\tilde g_{b\bar d}-
\tilde g_{a\bar d}\tilde g_{b\bar c}\right)\label{gflux}
\end{equation}
This anzatz solves the field equation $\nabla\cdot G=0$. The exterior derivative of the boundary
condition (\ref{cbc}) implies
\begin{equation}
G=\cases{
-{\epsilon\over \sqrt{2}}\left({\rm tr}(F\wedge F)-
\frac12{\rm tr}(R\wedge R)\right)
&on $M_1$\cr
+{\epsilon\over \sqrt{2}}\left({\rm tr}(F\wedge F)-
\frac12{\rm tr}(R\wedge R)\right)
&on $M_2$\cr}\label{grr}
\end{equation}
Since ${\rm tr}(R\wedge R)$ is a $(2,2)$ form, it takes a similar tensorial form to the flux term in
Eq. (\ref{gflux}).
The  boundary conditions are satisfied if ${\rm tr}(F\wedge F)={\rm tr}(R\wedge R)$ on the visible
brane $M_1$ and $F=0$ and on the hidden brane $M_2$. The value of $\alpha$ can be related through
Eq. (\ref{grr}) to an integer $\beta$ characterising the Pontrjagin class of the Calabi-Yau space
\cite{Lukas:1998tt},
\begin{equation}
\alpha={4\sqrt{2}\pi^2\over v_{CY}^{2/3}}\epsilon\beta
\end{equation}
where $v_{CY}$ is the volume of the Calabi-Yau space.

The volume function $V(z)$ is determined by the exact solution of the `$zz$'
component of the Einstein equations \footnote{
Our solution for $V$ is equivalent to the one used by Lukas et al.  in ref. \cite{Lukas:1998tt} when
adapted to our coordinate system. They express the solution as $V=b_0H^3$. It is also equivalent to
the background used by Curio and Krause in ref. \cite{Curio:2000dw}, $V=(1-{\cal S}_1 x^{11})^2$,
when their  ${\cal S}_1=\alpha V_1^{-2/3}/\sqrt{2}$.
},
\begin{equation}
V(z)=1-\sqrt{2}\alpha z.\label{v}
\end{equation}
The metric anzatz in consistent with all of the Einstein equations apart from the ones with
components along the Calabi-Yau direction, where the Einstein tensor vanishes but the stress energy
tensor is $O(\alpha^2)$. The difference between an exact solution to the Einstein equations and the
metric anzatz $\delta g_{IJ}=O(\alpha^2)$. If we calculate the action to reduce the theory
to four dimensions, then the error in the action is $O(\alpha^4)$. As long as we work within this
level of approximation we can use the Calabi-Yau approximation as the background for our reduced
theory. Note that this approximation is uniform in $z$, and having small values of $\alpha$ does not
necessarily mean small warping.

We shall also need to know the backgound solutions for a Rareta-Schwinger field when it takes
non-zero constant values on the boundary. It is conveniant to redefine the Rareta-Schwinger field
first by taking
\begin{equation}
\lambda_I=\Psi_I-\frac12\Gamma_I\Gamma^J\Psi_J\label{ldef}
\end{equation}
The Rareta-Schwinger equation for $\lambda_I$ becomes
\begin{equation}
\left(\Gamma^ID_I-\frac{\sqrt{2}}{96}\Gamma^{IJKL}G_{IJKL}\right)\lambda_P+
\frac{\sqrt{2}}{4}G_{PJKL}\Gamma^{JK}\lambda^L=0
\end{equation}
The solution with our metric/flux background and boundary conditions (\ref{gbc}) is
\begin{equation}
\lambda_\mu=V^{1/12}\theta_\mu\otimes u_++V^{1/12}\theta^*_\nu\otimes u_-\label{lb}
\end{equation}
where $u_\pm$ are the covariantly constant chiral spinors on the Calabi-Yau space and $\theta_\mu$
is a chiral $4-$spinor.

\section{reduction}

Reduction of low-energy heterotic $M-$theory to $5$ or $4$ dimensions follows a traditional route.
The light fields in the  $4-$dimensional theory correspond to the moduli of the background fields.
We shall be focussing especially on the volume of the Calabi-Yau space and the separation of the two
boundaries. These quantities can be expressed in terms of the values of $V$ on the two boundaries,
$V_1$ and $V_2$.

To allow for gravity in $4-$dimensions, the metric is replaced by
\begin{equation}
ds^2=V^{-2/3}dz^2+V^{-1/3}\Phi^2\tilde g_{\mu\nu}dx^\mu dx^\nu+
V^{1/3}(\tilde g_{a\bar b}dx^adx^{\bar b}+\tilde g_{\bar ab}dx^{\bar a}dx^b).\label{rmetric}
\end{equation}
The factor $\Phi^2$ is required to put the put the metric $\tilde g_{\mu\nu}$ into the Einstein
frame,
\begin{equation}
\Phi=\left(V_1^{4/3}-V_2^{4/3}\right)^{-1/2}.\label{phi}
\end{equation}
With this definition of the Einstein metric, the gravitational coupling in 4 dimensions is given by
\begin{equation}
\kappa_4^2={2\sqrt{2}\over 3}{\kappa^2_{11}\over v_{CY}}\alpha.
\end{equation}
 The reduction to the Einstein frame decouples the metric derivatives from $V_1$ and $V_2$. The
kinetic terms for $V_1$ and $V_2$ become
\footnote{For a detailed discussion see \cite{brax-2003-67}. The moduli fields used there are $Q$
and $R$, related to our variables by $V_1^{2/3}=Q\cosh R$ and $V_2^{2/3}=Q\sinh R$. Note that their
parameter $\alpha=\sqrt{3/2}$ for heterotic $M-$theory.} 
\begin{equation}
{1\over 2\kappa_4^2}\int_M \left( 
\frac32(\partial^2_{V_1}\ln\Phi)(\partial_\mu V_1)(\partial^\mu V_1)
+\frac32(\partial^2_{V_2}\ln\Phi)(\partial_\mu V_2)(\partial^\mu V_2)
+3(\partial_{V_1}\partial_{V_2}\ln\Phi)(\partial_\mu V_1)(\partial^\mu V_2)
\right)d\tau\label{v1v2}
\end{equation}
where $\tau$ is the volume element for the metric $\tilde g_{\mu\nu}$. Most authors like to
introduce an extra length-scale $\rho$ into the metric, which then appears in
the defintion of $\kappa_4$ \cite{lukas98,Lukas:1998tt,Lukas:1998ew}. However, this scale is a
redundant variable and gives the false impression that the brane separation can be chosen
arbitrarily.

For the Yang-Mills multiplets we have to rescale the gaugino to normalise the kinetic terms
correctly,
\begin{equation}
\chi_i=V^{1/4}\Phi^{-3/2}\tilde\chi_i\otimes u_++V^{1/4}\Phi^{-3/2}\tilde\chi_i^*\otimes u_-,
\end{equation}
where $u_\pm$ are the covariantly constant chiral spinors on the Calabi-Yau space.
The matter action for the hidden $E_8$ multiplet, for example, becomes
\begin{equation}
S_h={1\over 4 g^2}\int_{M}V_2
\left({\rm tr}(F_2^2)+\tilde\chi_2\gamma^{\mu}D_\mu\tilde\chi_2\right)d\tau.
\end{equation}
where the Yang-Mills coupling is
\begin{equation}
g^2={\kappa_{11}^2\over 2\epsilon v_{CY}}.
\end{equation}
Note that all of the model parameters $\kappa_{11}$, $\epsilon$, $\alpha$ and $v_{CY}$ can be
expressed in terms of the integer $\beta$ and measurable parameters $\kappa_4$ and $g$ for the
purposes of phenomenology.  

The reduction from 5 dimensions to 4 has also been done using a superfield formalism by Correia et
al \cite{Correia:2006pj}. This shows that in the $h_{1,1}=1$ case the reduced theory is a
supergravity model with $V_1$ and $V_2$ belonging to chiral superfields $S_1$ and $S_2$ with Kahler
potential
\begin{equation}
K=-3\ln\left((S_1+S_1^*)^{4/3}-(S_2+S_2^*)^{4/3}\right)
\end{equation}
Note that, for the real scalar components, the conformal factor introduced in Eq. (\ref{metric}) and
the Kahler potential are related by
\begin{equation}
\Phi=2^{2/3}e^{K/6},\label{phik}
\end{equation}
making the action derived from the K\"ahler potential consistent with Eq. (\ref{v1v2}). The
superfields $S_1$ and $S_2$ also appear in the reduced theory as gauge kinetic functions for the
$E_8$ Yang-Mills supermultiplets. 

\section{Condensates and fluxes}

Fermion condensates and fluxes of antisymmetric tensor fields may both play a role in the
stabilisation of moduli fields. In the context of low energy heterotic $M$-theory the most likely
candidate for forming a fermion condensate is the gaugino on the hidden brane, since the effective
gauge coupling on the hidden brane is larger and runs much more rapildy into a strong coupling
regime than the gauge coupling on the visible brane.

In the new formulation of low energy heterotic $M$-theory, gaugino condensates act as sources for
the field strength $G$ through the boundary conditions. Other non-zero contributions to the field
strength are possible, and we shall include some of these in the next section.

\subsection{Gaugino condensate}

The anzatz for a gaugino condensate on the boundary $M_i$ is \cite{Dine:1985rz},
\begin{equation}
\bar\chi_i\Gamma_{abc}\chi_i=\Lambda_i\omega_{abc}\label{cond}
\end{equation}
where $\Lambda_i$ depends only on the modulus $V_i$ and $\omega_{abc}$ is the covariantly constant
$3-$form on the Calabi-Yau space (i.e. the one with volume $v_{CY}$). The gaugino condensate
appears in the boundary conditions for the
antisymmetric tensor field and induces non-vanishing components
\begin{equation}
C_{abc}=\frac16\xi\omega_{abc}.
\end{equation}
where $\xi$ is a complex scalar field. The field strength associated with these tensor components is
\begin{equation}
G_{abcz}=-(\partial_z\xi)\omega_{abc}.
\end{equation}
When we solve the field equation $\nabla\cdot G=0$ with boundary conditions (\ref{cbc}), we get
\begin{equation}
\xi=-{\sqrt{2}\over 8}\Lambda_1\epsilon\Phi^2\left(V^{4/3}-V_2^{4/3}\right)
-{\sqrt{2}\over 8}\Lambda_2\epsilon\Phi^2\left(V^{4/3}-V_1^{4/3}\right).
\end{equation}
where $\Phi$ was defined in eq. (\ref{phi}).
The new flux contribution is
\begin{equation}
G_{abcz}=-\frac{\alpha}3(\Lambda_1+\Lambda_2)
\epsilon\Phi^2\omega_{abc}V^{1/3}.\label{gabc}
\end{equation}
The appearance of $\Phi$ in this formula is independent of the normalisation of the metric, but its
presence here will prove essential for the consistency of the reduction.

The non-zero flux depends on $V_1$ and $V_2$ through the $\Phi$ term and through $\Lambda$, which
depends on the volume factors $V_1$ and $V_2$ in the gaugino couplings. The $G^2$ term in the
action reduces to a potential $V_c$ in the Einstein frame, where
\begin{equation}
V_c=\frac{\sqrt{2}}3{\epsilon^2 v_{CY}\over 2\kappa_{11}^2}\alpha \Phi^6
|\Lambda_1+\Lambda_2|^2.\label{vc}
\end{equation}
This is not the whole story, however, and we shall attempt to find the potential by a better method,
using a reduction of the fermion sector in Sect. \ref{sp}.  

\subsection{Fluxes}

The boundary conditions on the flux $G$ do not fully determine the value of $G_{abcz}$, and it is
possible to have a non-zero flux of the form (\ref{gabc}) even in the absence of a gaugino
condensate. In this situation the main restriction is topological, related to the quantisation rule
\cite{Rohm:1985jv,Witten:1996md},
\begin{equation}
{1\over4\pi \epsilon} \int_{C_4} {G\over 2\pi}+{1\over32\pi^2}\int_{C_4}R\wedge R=n
\end{equation} 
where $C_4$ is a any closed four-cycle. We have to apply this rule in the presence of boundaries, 
where there are modifications as suggested by Lukas et al.  \cite{Lukas:1997rb}. We shall use,
\begin{equation}
{1\over4\pi \epsilon} \left(\int_{C_4} G-6\int_{\partial C_4}C_{\partial M}\right)
+{1\over32\pi^2}\int_{C_4}R\wedge R=n,\label{qg}
\end{equation}
where $C_{\partial M}$ denotes the restriction of the antisymetric tensor $C$ to the boundary.

We modify our earlier anzatz so that
\begin{eqnarray}
C_{abc}&=&\frac16\xi\omega_{abc},\\
C_{zab}&=&\frac1{12}(\partial_z\sigma)\,b_{ab}.\label{cz}
\end{eqnarray}
This anzatz introduces a non-trivial topological structure with locally defined antisymmetric tensor
components $b_{ab}$, defined such that $db=\omega$. These components can be related to a gerbe, as
described in the appendix. Note that the derivative in Eq. (\ref{cz}) is needed to ensure that $C$
is the $3-$form associated with a $2-$gerbe.

The solution to the field equation $\nabla\cdot G=0$ is now
\begin{equation}
G_{abcz}=-\frac{\alpha}3(\Lambda_1+\Lambda_2+\lambda)
\epsilon\Phi^2\omega_{abc}V^{1/3}\label{gnew}
\end{equation}
where $\lambda$ depends on the values of $\sigma$ on the boundaries $M_1$ and $M_2$,
\begin{equation}
\lambda={4\sqrt{2}\over\epsilon}(\sigma_1-\sigma_2)
\end{equation}
The $\lambda$ term is the $M$-theory analogue of the flux term introduced by Dine et al.
\cite{Dine:1985rz} in an attempt to cancell the gaugino condensate in type IIB superstring theory.

The value of $\lambda$ can be determined by the quantisation rule (\ref{qg}) using 
$C_4=S_1/Z_2\times C_3$,
\begin{equation}
\lambda={32\sqrt{2}\pi^2\over c v_{CY}^{1/2}}n,\qquad n=0,1,\dots
\end{equation}
where the constant $c$ is defined by
\begin{equation}
c={1\over v_{CY}^{1/2}}\int_{C_3}\omega.
\end{equation}
This is the analogue of the flux quantisation in type IIB superstring theory \cite{Rohm:1985jv}. We
can also obtain the quantisation rule by calculating the gerbe charge of $\sigma$, using the rules
described in appendix.

\subsection{Superpotential}\label{sp}

The superpotential $W$ for the moduli superfields $S_1$ and $S_1$ can be obtained from the gravitino
mass terms
\begin{equation}
{\cal L}_{3/2\hbox{ mass}}={1\over 2\kappa_4^2}
e^{K/2}\left(W\bar\theta^\mu\theta_\mu+c.c\right)\label{defw}
\end{equation}
provided that the $\theta_\mu$ are correctly normalised,
\begin{equation}
{\cal L}_{3/2\hbox{ kinetic}}=
{1\over 2\kappa_4^2}\left(\bar\theta^\mu\gamma^\nu D_\nu\theta_\mu+c.c\right).
\label{defk}
\end{equation}
Note that these are valid in the Einstein frame and indices are raised with the metric 
$\tilde g^{\mu\nu}$.

The $4-$dimensional action for the gravitino field can be obtained using the background (\ref{lb})
and metric (\ref{rmetric}) 
\footnote{The condensate will change the gravitino background to order $\alpha\Lambda$
 and the action to order 
$(\alpha\Lambda)^2$. We are therefore entitled to ignore this correction to the background when
calculating the superpotential.}
.  We introduce an additional factor $a\equiv a(V_1,V_2)$ to allow us to
adjust the normalisation, 
\begin{equation}
\lambda_\mu=aV^{1/12}\theta_\mu\otimes u_++a^*V^{1/12}\theta^*_\nu\otimes u_-
\end{equation}
For the kinetic term,
\begin{equation}
\bar\lambda^\mu\Gamma^\nu D_\nu\lambda_\mu|g|^{1/2}=
|a|^2\Phi\,V^{2/3}
\left(\bar\theta^\mu\gamma^\nu D_\nu\theta_\mu+c.c.\right)|\tilde g|^{1/2}.\label{rskin}
\end{equation}
Similarly, for the flux term,
\begin{equation}
\bar\lambda^\mu\Gamma^{abcz}G_{abcz}\lambda_\mu|g|^{1/2}=
|a|^2\Phi^2\,V^{1/3}\omega^{abc}G_{abcz}
\left(\bar\theta^\mu\theta_\mu+c.c.\right)|\tilde g|^{1/2}.\label{rsflux}
\end{equation}
These terms have to be integrated over the Calabi-Yau space and over the $z$ coordinate. However,
when we substitute the flux from eq. (\ref{gnew}), both terms have an identical dependence on $z$.
Therefore the superpotential can be read off directly from eqs. (\ref{rskin}) and (\ref{rsflux}) by
comparison with eqs. (\ref{defk}) and (\ref{defw}). The
normalisation factor $\Phi$ cancells due to eq. (\ref{phik}), and we get
\begin{equation}
W=-3\sqrt{2}\alpha \epsilon\,(\Lambda_1+\Lambda_2+\lambda)\label{superpot}
\end{equation}
where $\Lambda_1$ and $\Lambda_2$ are the amplitudes of the condensates (\ref{cond}) and $\lambda$
is the quantised flux.

Most discussions of the condensate induced superpotential do not take the warping of the metric into
account.  We have found that the warping of the metric background has had no effect on the
superpotential. Krause \cite{Krause:2007gj} also finds that the warping does not affect the
condensate contribution to the superpotential, but he claims a warping dependence in the flux term.
This can be traced to a formula for the superpotential given by Anguelova and Zoubos
\cite{Anguelova:2006qf}. We can derive a similar formula by integrating Eq. (\ref{rsflux}) and
comparing to Eq. (\ref{defw}),
\begin{equation}
W={4\alpha\over  v_{CY}}{|a|^2\over \Phi}
\int_{Y\times[z_1,z_2]}\,V^{-1/6}G\wedge V^{1/2}\bar\omega.
\end{equation}
Note that $a$ depends on the moduli fields when we normalise the gravitino kinetic term using
(\ref{rskin}). The result derived by Anguelova and Zoubos does not contain the factor 
$|a|^2/\Phi$ \footnote{If we use their gravitino field $\propto V^{-1/12}$ rather than our solution 
$\propto V^{1/12}$, the factor disappears but the $V^{-1/6}$ inside the integral becomes $V^{-1/2}$.
The final superpotential obtained after integration is unchanged.}
.

\section{Moduli stabilisation}

Moduli stabilisation can be achieved by following a similar patern to moduli stabilisation in type
IIB string theory \cite{Kachru:2003aw}. The first stage involves finding a suitable superpotential
which fixes the moduli but leads to an Anti-de Sitter vaccum. The negative energy of the vaccum
state is then raised by adding a non-supersymmetric contribution to the energy.

The potential is given in terms of the Kahler potential $K$ and the superpotential $W$,
\begin{equation}
V=\kappa_4^{-2}e^K\left(
g^{i\bar \jmath}(D_iW)(D_j W)^*-3WW^*
\right),
\end{equation} 
where $g_{i\bar\jmath}$ is the hessian of $K$ and
\begin{equation}
D_iW=e^{-K}\partial_{V_i}(e^K\,W).
\end{equation}
Minima of the potential occur when $D_iW=0$. If these minima exist, their location is fixed under
supersymmetry transformations. However, the boundary conditions at the potential minima are not
generally preserved by supersymmetry and the theory at a supersymmetric minimum is not necessarily
supersymmetric. This distinction is subtle, but important because it allows for mechanisms which
produce de Sitter minima.

We shall examine the supersymmetric minima of the potential for two toy models. We shall concentrate
on general features rather than obtaining a precise fit with particle phenomenology.

\subsection{Double-condensate}

Following the type IIB route, we assume the existence of a flux term $W_f$ in the superpotential
which stabilises the $(2,1)$ moduli, and then remains largely inert whilst the other moduli are
stabilised.

The gauge coupling on the hidden brane runs to large values at moderate energies and this is usually
taken to be indicative of the formation of a gaugino condensate. Local supersymmetry restricts the
form of this condensate to \cite{Burgess:1995aa}
\begin{equation}
\Lambda_2=B_2 \,v_{CY}^{-1/2}e^{-\mu V_2}
\end{equation}
where $B_2$ is a constant and $\mu$ is related to the renormalisation group $\beta$-function by
\begin{equation}
\mu={6\pi\over b_0\alpha_{GUT}},\quad \beta(g)=-{b_0\over 16\pi^2}g^3+\dots.
\end{equation}
The gauge coupling on the visible brane is supposed to run to large values only at low energies to
solve the hierarchy problem, and a low energy condensate would have a negligable effect on moduli
stabilisation. There might, however,  be a separate gauge coupling from part of the $E_6$ symmetry
on the visible brane which becomes large at moderate energies with a significant condensate term.
The requirement for this to happen is a large $\beta$-function, possibly arising from charged
scalar field contributions. The total superpotential for such a model would be given by combining
Eq. (\ref{superpot}) with
$W_f$,
\begin{equation}
W=be^{-\mu V_2}+ce^{-\tau V_1}-w,
\end{equation}
where $w=-W_f$ and $b$, $c$ are constants, which we assume to be real but not necessarily positive.

The fields at the minimum of the potential could be complex, and we therefore separate real and
imaginary parts,
\begin{equation}
V_i=u_i+i v_i.
\end{equation}
The superderivatives of the potential are
\begin{eqnarray}
D_1W&=&-c\tau e^{-\tau V_1}-2(u_1^{4/3}-u_2^{4/3})^{-1}u_1^{1/3}W,\\
D_2W&=&-b\mu e^{-\mu V_2}-2(u_1^{4/3}+u_2^{4/3})^{-1}u_2^{1/3}W.
\end{eqnarray}
Solving for the values of $V_1$ and $V_2$ at the minimum of the potential is not very informative.
Instead, we express the parameters $b$, $c$ and $d$ in terms of the values of $V_1$ and $V_2$ at
the supersymmetric minimum,
\begin{eqnarray}
{b\over w}&=&
{-2u_2^{1/3}e^{\mu V_2}\mu^{-1}\over u_1^{4/3}-u_2^{4/3}-2\mu^{-1}u_2^{1/3}+2\tau^{-1}u_1^{1/3}},\\
{c\over w}&=&
{2u_1^{1/3}e^{\tau V_1}\tau^{-1}\over u_1^{4/3}-u_2^{4/3}-2\mu^{-1}u_2^{1/3}+2\tau^{-1}u_1^{1/3}}.
\end{eqnarray}
We conclude from these expressions that, if $b/w$ and $c/w$ are real, then $V_1$ and $V_2$ are both
real. (If $b$ and $c$ are not real, then it becomes difficult to satisfy the background field
equations on the antisymmetric tensor field with the resulting complex boundary conditions).

\begin{center}
\begin{figure}[ht]
\scalebox{1.0}{\includegraphics{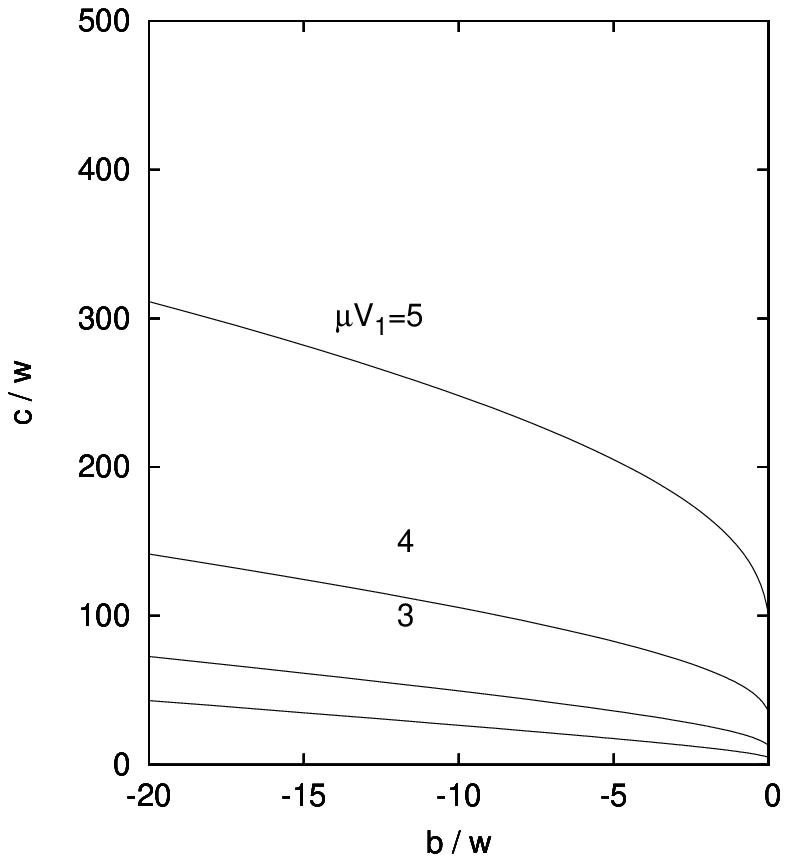}}
\scalebox{1.0}{\includegraphics{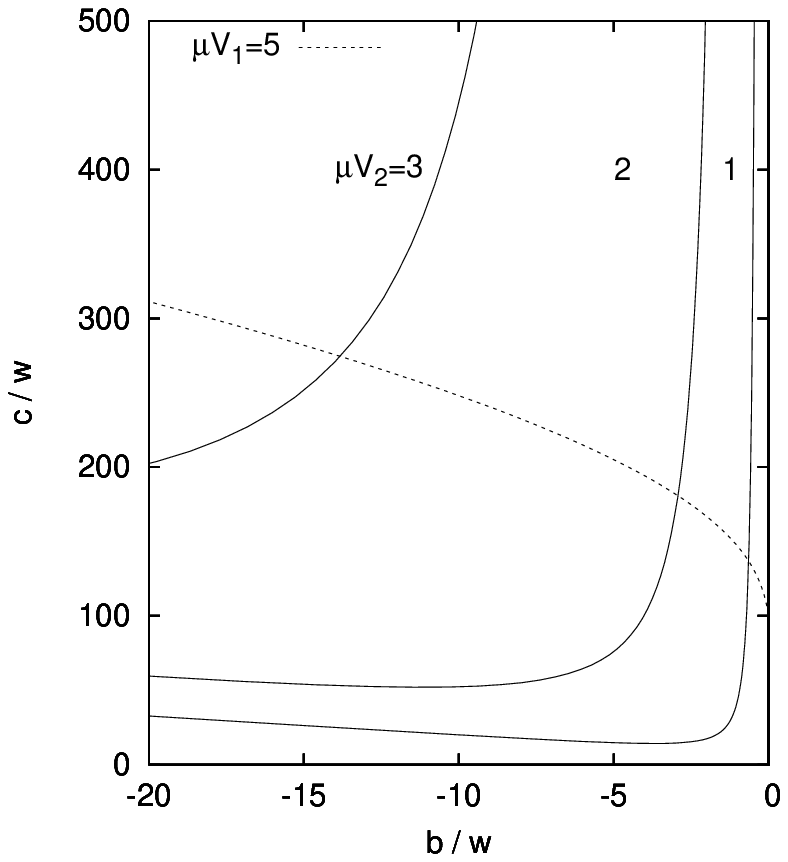}}
\caption{The values of the volume moduli at the minimum of the potential with two condensates and
$\tau/\mu=1.2$. The left panel shows values of $V_1$ and the right panel shows values of $V_2$. }
\label{contours2g}
\end{figure}
\end{center}

Supersymmetric minima exist for $b<0$ and $c>0$. The values of $V_1$ and $V_2$ at the minima are
shown in Fig. \ref{contours2g}.
At the minima of the potential, the flux term $|W_f|$ is larger than the gauge condensate terms.
This is consistent with the idea that we consider the stabilisation of the $(2,1)$ moduli
independently of the other moduli.

\subsection{Other non-perturbative terms}

If there are no high energy condensates on the visible brane, then we can replace the condensate
with another non-perturbative effect. The usual candidate for this is a membrane which
stretches between the two boundaries. The area of the membrane $\propto V_1-V_2$ and the type of
contribution this gives to the superpotential is
\begin{equation}
W_{np}=ce^{-\tau(V_1-V_2)}.
\end{equation}
The total superpotential for the toy model is given by
\begin{equation}
W=be^{-\mu V_2}+ce^{-\tau(V_1-V_2)}-w,
\end{equation}
where $w=-W_f$ and $b$, $c$ are constants. . 

\begin{center}
\begin{figure}[ht]
\scalebox{1.0}{\includegraphics{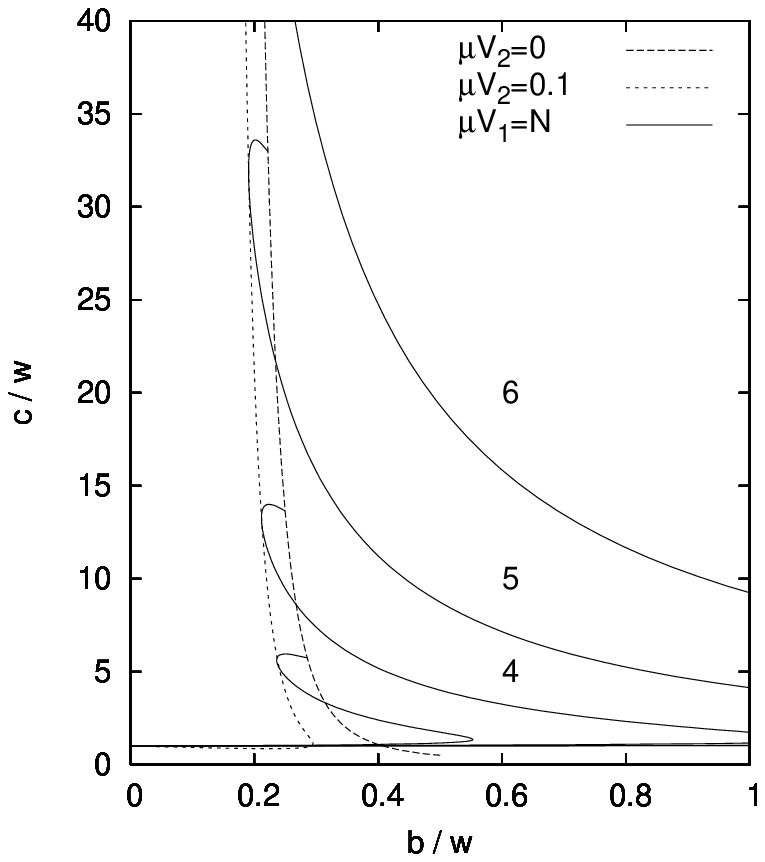}}
\scalebox{1.0}{\includegraphics{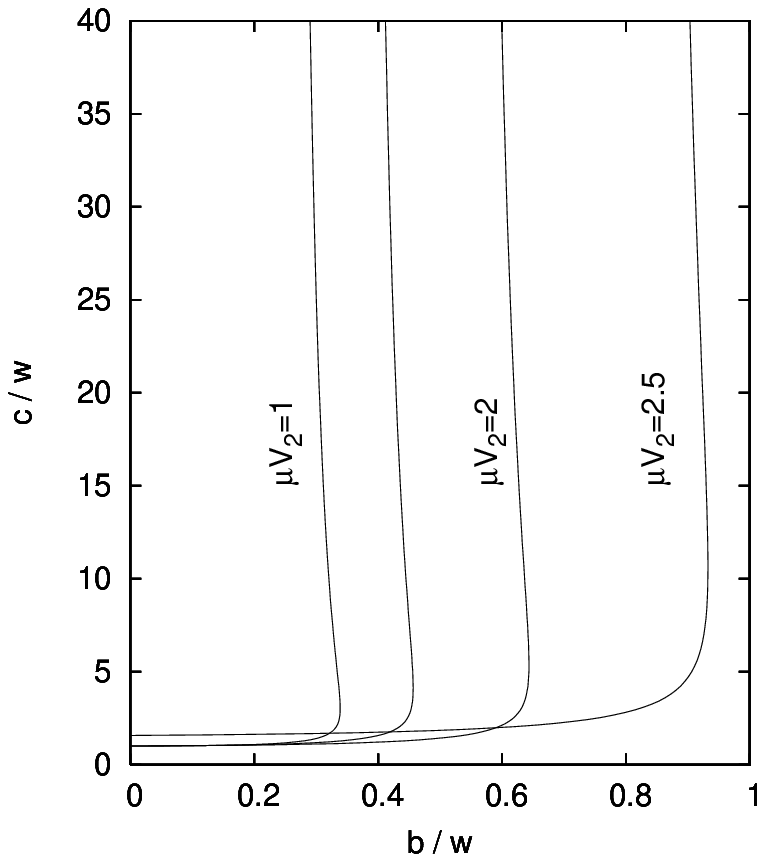}}
\caption{The values of the volume moduli at the minima of the potential with $W_{np}$ and
$\mu=\tau$. The left
panel shows values of $V_1$ and the right panel shows values of $V_2$. There are no solutions to
the left of the leftmost dashed line ($\mu V_2=0.1$) and one solution to the right of the rightmost
dashed line. The strip between the dashed lines indicates a parameter region with small values of
$V_2$. }
\label{contours}
\end{figure}
\end{center}

This time the parameters $b$, $c$ and $d$ given in terms of the values of $V_1$ and $V_2$ at
the supersymmetric minimum are
\begin{eqnarray}
{b\over w}&=&
{-2u_2^{1/3}e^{\mu V_2}\mu^{-1}\over 
u_1^{4/3}-u_2^{4/3}+2\mu^{-1}(u_1^{1/3}-u_2^{1/3})+2\tau^{-1}u_1^{1/3}},\\
{c\over w}&=&
{2u_1^{1/3}e^{\tau V_1}\tau^{-1}\over 
u_1^{4/3}-u_2^{4/3}+2\mu^{-1}(u_1^{1/3}-u_2^{1/3})+2\tau^{-1}u_1^{1/3}}.
\end{eqnarray}
Again we conclude from these expressions that $V_1$ and $V_2$ are both real.
The values of the moduli at the supersymmetric minima of the potential are shown in figure
\ref{contours}, where we have taken $\tau=\mu$. Other values of $\tau$ give a qualitatively similar
figure. There are always supersymmetric minima in the parameter region indicated on the figure. 

\section{conclusion}

We have started from an improved formulation of low energy heterotic $M$-theory and revisited some
aspects of gaugino condensation and moduli stabilisation. Many of the undesirable features that
have been introduced previously, such as hiding away delta functions in field redefinitions and
modifying the Bianchi identities have now become unnecessary. The reduction of low energy heterotic
$M$-theory to four dimensions also displays some new features. The non-trivial topology of the
anti-symmetric tensor field shows up very clearly, for example.

Having an action which is valid to all orders in the gravitational coupling means than now the
warping of the five-dimensional metric can be taken into account consistently. We would like to
stress that a small brane charge $\alpha$ does not necessarily imply small warping.

The final superpotential contains no surprises. It takes the standard form expected for a gaugino
condensate in any supersymmetric theory \cite{Burgess:1995aa}, and both condensate and quantised
flux contributions to the superpotential contain no corrections due to the warping of the metric in
higher dimensions. 

It remains to be seen how the other ingredients of low
energy heterotic $M$-theory which we have neglected in this paper enter into the mix, for example
the extra $(1,1)$ moduli, five-branes and anti five-branes may all play a role in a realistic model
\cite{Gray:2007qy}. Some features of the present calculation may be helpful in these
generalisations. For example, the five dimensional superfield formalism gives a good guide as to
good choices of moduli fields, in our case these where the Calabi-Yau volumes rather than the brane
separation. The inclusion of five-branes in the improved formalism for heterotic $M$-theory still
remains to be developed. 

We have not made full use in this paper of the fermion boundary conditions (see Eq. (\ref{gbc})).
These will break the supersymmetry in the presence of the gauge condensate and a non-zero quantum
vacuum energy will result. We are presently considering situations where this vacuum energy can
raise the vacuum energy at the minimum of the potential to positive values. 

\acknowledgements
We are grateful to Lilia Anguelova and Konstantinos Zoubos for discussing their work and to 
Ezra Getzler for illuminating discussions about gerbes.

\appendix

\section{gerbe connections}

In this appendix we give a simplified description of connections on gerbes, following the review by
Hitchin \cite{Hitchin:1999fh}. We shall define a $1-$gerbe with respect to a given open
cover $U_\alpha$ of a manifold. We use the notation $U_{\alpha\beta\dots\gamma}$ to refer to the
intersection of the sets $U_\alpha$ \dots $U_\gamma$. The same subscripts on a tensor indicate that
the tensor is only defined on the corresponding region. 

A connection with charge $q$ on a $1-$gerbe is defined to be a set
\begin{equation}
\{B_\alpha,A_{\alpha\beta},g_{\alpha\beta\gamma}\}\label{gerbe}
\end{equation}
where the $B_\alpha$ are $2-$forms,  $A_{\alpha\beta}$ are $1-$forms and  $g_{\alpha\beta\gamma}$
are complex numbers of unit modulus. The forms are related by a set of consistency relations,
\begin{eqnarray}
B_\alpha-B_\beta&=&-dA_{\alpha\beta}\\
A_{\alpha\beta}+A_{\beta\gamma}+A_{\gamma\alpha}&=&-i\,q^{-1}\,d \ln\,g_{\alpha\beta\gamma}\\
g_{\alpha\beta\gamma}g_{\gamma\beta\delta}
g_{\beta\gamma\delta}g_{\alpha\gamma\delta}&=&1.
\end{eqnarray}
A simple generalisation allows the definition of $p-$gerbes with connection for $p=0,1\dots$, the
leading terms being $(p+1)-$forms. The case $0-$gerbe with connection is equivalent to a $U(1)$
fibre bundle with connection.

The curvature of the $p-$gerbe with connection is defined to be $d B_\alpha$ and it is independent
of the choice of open set $\alpha$. It defines an integral class, i.e.
\begin{equation}
\int_{C_{p+2}} {dB\over 2\pi}={n\over q},
\end{equation}
where $n$ is an integer and $C_{p+2}$ is a closed cycle. This generalises the Dirac quantisation
condition to $p-$forms. The converse also holds, i.e. given a closed $p+2$ form and the
quantisation condition, then there exists a gerbe and connection.
In the case of supergravity, it is the existence of a quantisation condition for the antisymmetric
tensor flux which indicates that the antisymmetric field should be associated with a gerbe.

The wedge product of two connections on a gerbe, lets say $B$ and $C$, defined by taking the wedge
products of the form components term by term, usually fails to satisfy the consistency relations.
However, with a little care, it is possible to define $B\wedge dC$ and $C\wedge dB$ so that these
are gerbe connections. If $B$ is a connection on a $p_1-$gerbe with charge $q_1$ and $C$  is a
connection on a $p_2-$gerbe with charge $q_2$, then
$B\wedge dC$ and $C\wedge dB$
are connections on a $p_1+p_2+1-$gerbe with charge $q_1q_2$.

We can apply the gerbe technology to the Calabi-Yau $3-$form $\omega$. Introduce a closed cycle
$C_3$ and define a constant $c$ by
\begin{equation}
\int_{C_3}\omega=c\,v_{CY}^{1/2}
\end{equation}
There is only one 3-cycle on the Calabi-Yau space which gives a non-vanishing result, and the
associated quantisation condition implies there exists a gerbe connection $b$ with curvature
$db=\omega$ and charge
$q_b=2\pi c^{-1}v_{CY}^{-1/2}$. 
\acknowledgements

\bibliography{paper.bib}

\end{document}